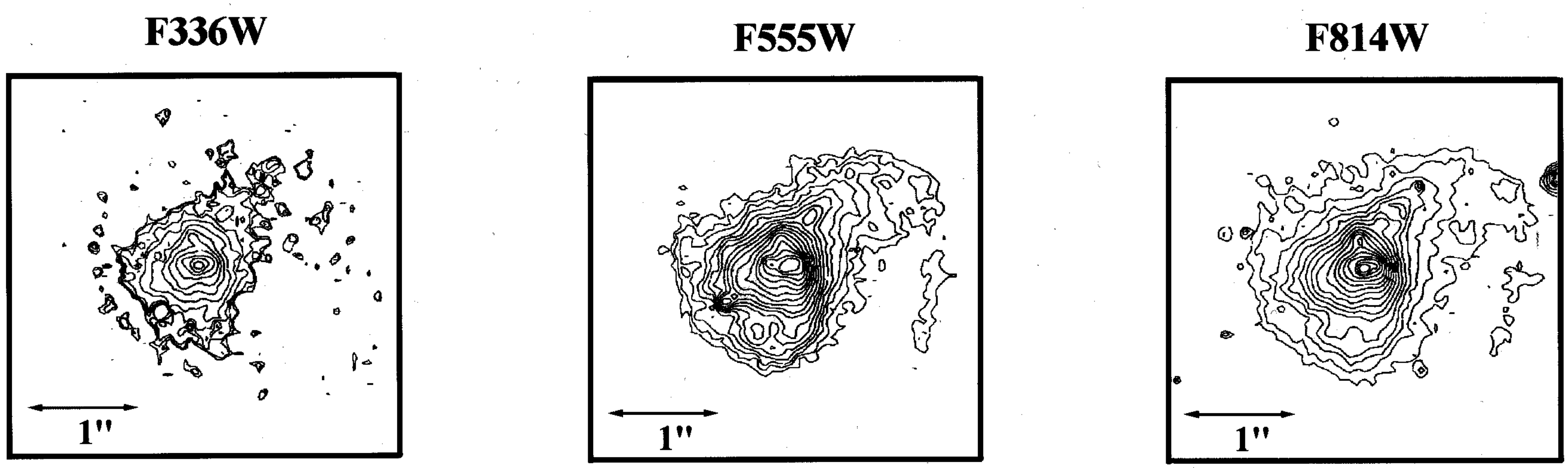

F336W
1"
F555W
1"
F814W
1"

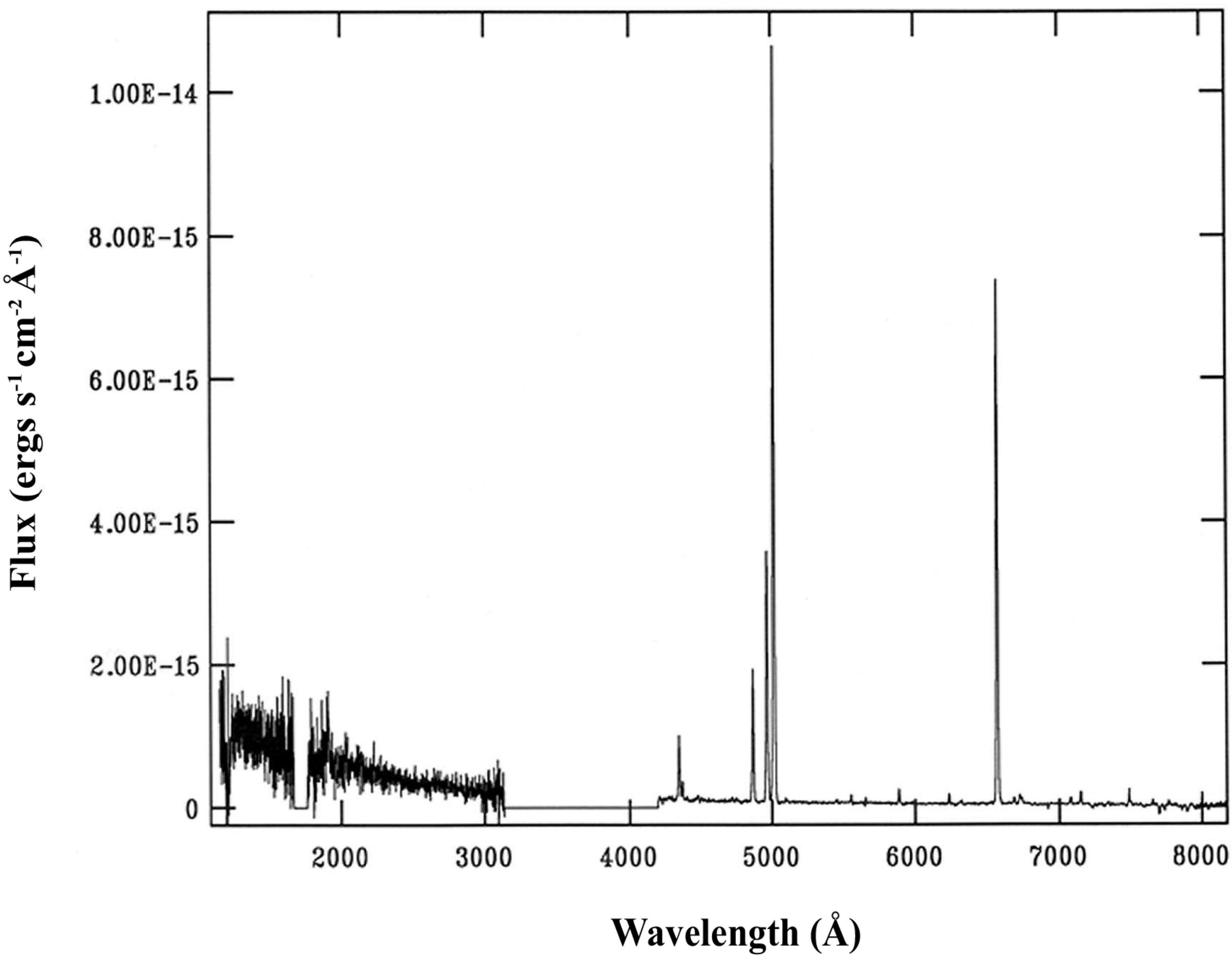

Flux (ergs s-1 cm-2 Å-1)
1.00E-14
8.00E-15
6.00E-15
4.00E-15
2.00E-15
0
2000
3000
4000
5000
6000
7000
8000
Wavelength (Å)

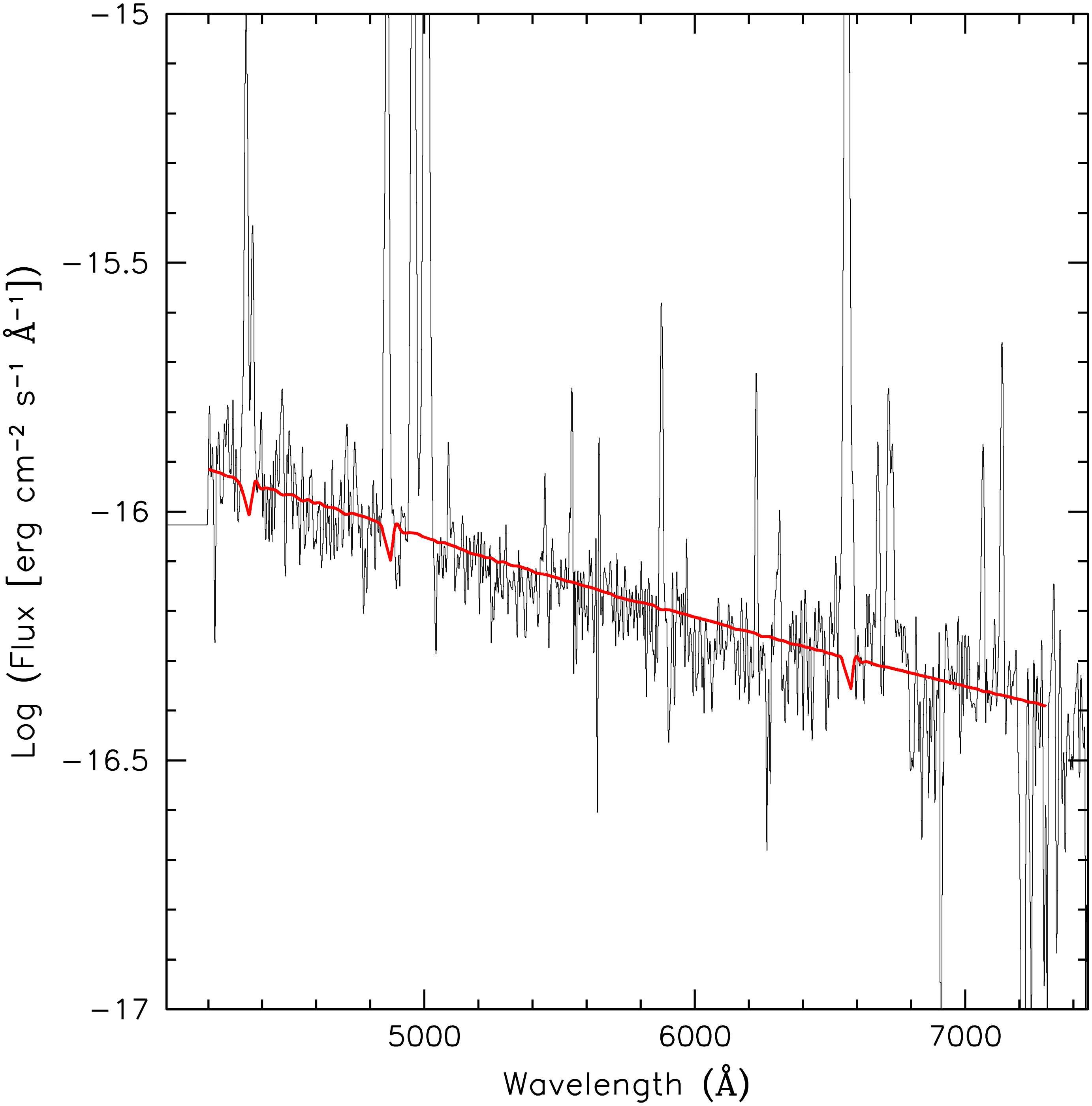

−15
−15.5
−16
−16.5
−17
5000
6000
7000
Log (Flux [erg cm−2 s−1 Å−1])
Wavelength (Å)

# POX 186: A Dwarf Galaxy in the Process of Formation?[1]

Michael R. Corbin

Science Programs, Computer Sciences Corporation

Space Telescope Science Institute, 3700 San Martin Drive

Baltimore, MD 21218; corbin@stsci.edu

&

William D. Vacca

Max-Planck Institut für Extraterrestrische Physik

Giessenbachstrasse Postfach 1312, Garching, D-85741, Germany;

vacca@mpe.mpg.de

## ABSTRACT

We present deep *U*, *V* and *I* band images of the "ultracompact" blue dwarf galaxy POX 186 obtained with the Planetary Camera 2 of the *Hubble Space Telescope*. We have also obtained a near-ultraviolet spectrum of the object with the Space Telescope Imaging Spectrograph, and combine this with a new ground-based optical spectrum. The images confirm the galaxy to be extremely small, with a maximum extent of only 300 pc, a luminosity $\sim 10^{-4}\ L^*$ and an estimated mass $\sim 10^7\ M_{\odot}$. Its morphology is highly asymmetric, with a tail of material on its western side that may be tidal in origin. The *U*-band image shows this tail to be part of a stream of material in which stars have recently formed. Most of the star formation in the galaxy is however concentrated in a central, compact ($d \sim 10 – 15$ pc) star cluster. We estimate this cluster to have a total mass $\sim 10^5\ M_{\odot}$, to be forming stars at a rate of $< 0.05\ \mathrm{yr}^{-1}$, and to have a maximum age of a few million years. The outer regions of the galaxy are significantly redder than the cluster, with $V – I$ colors consistent with a population dominated by K and M stars. From our analysis of the optical spectrum we find the galaxy to have a metallicity $Z \cong 0.06\ Z_{\odot}$ and to contain a significant amount of internal dust ($E(B – V) \cong 0.28$); both values agree with previous estimates. While these results rule out earlier speculation that POX 186 is a protogalaxy, its morphology, mass and active star formation suggest that it represents a recent (within $\sim 10^8$ yr) collision between two clumps of stars of *sub-galactic* size (~100 pc). POX 186 may thus be a very small dwarf galaxy that, dynamically speaking, is still in the process of formation. This interpretation is supported by the fact that it resides in a void, so its morphology cannot be explained as the result of an encounter with a more massive galaxy. Clumps of stars this small may represent the building blocks required by hierarchical models of galaxy formation, and these results also support the recent "downsizing" picture of galaxy formation in which the least massive objects are the last to form.

# 1. INTRODUCTION

The blue compact dwarf galaxy (BCDG) POX 186 (= PGC 046982) is an intriguing object. Kunth, Sargent & Kowal (1981) first identified it as a nearby emission-line galaxy with a heliocentric radial velocity of 1170 km $s^{-1}$. Ground-based CCD images obtained by Kunth, Maurogordato & Vigroux (1988) revealed it to be very faint and compact, with $m_p$ = 17 and an angular extent of only about 3″. In contrast to the other BCDGs imaged in their study, Kunth et al. (1988) found no clear evidence of an underlying population of evolved stars surrounding the star-forming regions of POX 186, and speculated that the object is a protogalaxy, forming its first generation of stars. The spectroscopy of Kunth & Sargent (1983) however revealed metal lines in the galaxy's spectrum, as well as evidence of internal dust. This suggests earlier episodes of star formation, and argues against the protogalaxy hypothesis.

More recently, Doublier et al. (2000; hereafter D2000) obtained deep images of POX 186 from the ground under sub-arcsecond seeing conditions. The D2000 images confirm the galaxy to be "ultracompact," with a maximum projected size of less than a kiloparsec at its redshift. It is thus one of the smallest galaxies known. D2000 also find a clear $R – I$ color gradient in the object, with the outer regions being significantly redder ($R – I \cong 0.4$) than the galaxy's center ($R – I \cong -1.0$). This supports the spectroscopic evidence that previous episodes of star formation have occurred, and the absence of early-type stars in these outer regions implies a lower limit of ~$10^8$ years to the age of the underlying stellar population of the galaxy if standard initial mass functions are assumed. In this respect POX 186 is similar to larger BCDGs, the majority of which consist of clusters of early-type stars embedded in envelopes of redder stars similar in size and shape to normal dwarf elliptical galaxies (Loose & Thuan 1985; Kunth et al. 1988; Papaderos et al. 1996).

However, in contrast to larger BCDGs, the ground-based images of Kunth et al. (1988) and D2000 indicate that POX 186 has an asymmetric morphology. Kunth et al. (1988) note the presence of an "arm" of material on the galaxy's western side, which is also suggested in the D2000 images. In addition to its small size, this morphology suggests that POX 186 is an extreme member of the BCDG class. Detailed study of it may thus offer more general insight into the phenomenology of these objects. In particular, the very small angular size of the galaxy makes it a good target for *Hubble Space Telescope* (HST) observations, as it is difficult to resolve from the ground.

With these motivations we obtained deep *U*, *V*, and *I* images of POX 186 with the high resolution Planetary Camera 2 (PC2) of HST, along with an ultraviolet spectrum of the object obtained with the Space Telescope Imaging Spectrograph (STIS). We combine this spectrum with a new ground-based optical spectrum for an analysis of the galaxy's metallicity and stellar population. Our principal finding is that the asymmetric morphology of POX 186 as well as its active star formation may be the result of a recent collision between two clumps of stars of *sub-galactic* size (~100 pc). This in turn suggests that POX 186 is a very small dwarf galaxy that, at least in terms of its stellar dynamics, is still in the process of formation.

In the following section we describe our observations. We present their analysis in § 3, and conclude with a discussion in § 4. A cosmology of $H_0$ = 72 km $s^{-1}$ $Mpc^{-1}$, $\Omega_M$ = 0.3, $\Omega_\Lambda$ = 0.7 is assumed.

## 2. OBSERVATIONS

Our HST observations of POX 186 are summarized in Table 1. The PC2 images were taken in the F336W, F555W and F814W filters, which are similar to the Johnson *U*, *V* and *I* filters. These filters were selected to cover the widest possible wavelength range allowed by the PC2, and to distinguish the emission from the early-type and late-type stars in the galaxy. Integration times were chosen to allow the detection of individual OB and M-giant stars in the F336W and F814W filters, respectively, at the distance to the galaxy under our adopted cosmology (21 Mpc). STIS observations were made with the MAMA detectors, using the G140L and G230L gratings, which cover a combined wavelength range of approximately 1200 Å – 3100 Å. The slit used was 52″ × 0.5″. Integration times were determined from the magnitudes of Kunth et al. (1988) and the assumption that the spectral shape of the galaxy in the near ultraviolet is similar to that of the O stars producing its strong nebular emission lines. As discussed in the next section, we find most of the star formation in POX 186 to be concentrated in a marginally resolved star cluster near the geometric center of the galaxy. The resolutions of the STIS spectra should thus be close to those for a point source, which we assume is centered in the slit. These resolutions are 1.5 Å at 1500 Å for G140L, and 2.1 Å at 2400 Å for G230L for this aperture. The spectra were extracted from an aperture of 0.26″.

We use the "On-The-Fly" calibrated versions of the images and spectra produced by the HST calibration pipeline for our analysis, after additional cosmic ray removal. The PC2 images were flux calibrated using the conversion factors in the image headers. The calibrated images and spectra were also corrected for Galactic extinction using the extinction values listed in the NASA Extragalactic Database, which are based on $A_B$ = 0.201 (Schlegel, Finkbeiner & Davis 1998) and the extinction law of Cardelli, Clayton & Mathis (1989).

An optical spectrum of POX 186 was obtained on UT 1999 March 13 using the 90" Bok telescope of Steward Observatory, over the spectral range 4200 Å - 8150 Å at a spectral resolution of 6 Å. Conditions were photometric, seeing was ~1.2″, and the resulting spectrum was flux calibrated using standard star spectra taken before and after the integrations. The total integration time was two hours, the slit width was 4.5″, and the slit position angle was 0°. The airmass range was 1.38 to 1.41. Under these conditions the whole galaxy should fit within the slit, and the resulting flux calibration should be accurate to within ~10% - 20%. From the emission lines in this spectrum we measure the redshift of POX 186 relative to the Local Standard of Rest to be 0.00398 ± 0.00001. This agrees with the original measurement of Kunth et al. (1981). Correcting this redshift to the frame of the cosmic microwave background (Fixsen et al. 1996) yields a distance of 21 Mpc to the galaxy under our adopted cosmology.

## 3. RESULTS AND ANALYSIS

### 3.1 Images

A color composite of the F336W, F555Wand F814W images is shown in Figure 1. The nominal angular resolution of the PC2 chip of ~0.1″ corresponds to a spatial scale ~10 pc at our adopted distance.

The asymmetry seen in POX 186 in the Kunth et al. (1988) and D2000 images is resolved into a narrow tail of stars that bends back towards the main body of the galaxy, although there appear to be stars or groups of stars in the intervening space. In addition to the greater detail seen in POX 186, this image reveals several faint red objects near it, particularly to the northeast. Close examination of the morphologies and brightness profiles of these objects suggests that they are background galaxies, possibly a group. The object approximately 5″ southwest of POX 186 in particular appears to be a face-on spiral galaxy, rather than a group of stars associated with POX 186. The unresolved source at the northwestern tip of POX 186 is also only seen in the F814W image, and could be a faint Galactic late-type star, as there is no evidence of an extended structure or of an association with POX 186.

Contour plots of the individual images are shown in Figure 2. These plots, particularly for the F336W and F814W images, show the galaxy’s emission to be dominated by a single peak near its geometric center. This contrasts with the result of D2000, who claim evidence of three peaks of comparable brightness on the basis of the deconvolution of their ground-based *R* and *I* images. Secondary peaks are seen in our images (particularly F555W) to the north and southeast of the central peak, and may coincide at some level with the regions identified by D2000. However, in our images the flux of these regions is only ~1% - 5% of that of the central brightness peak.

We attempted to deconvolve our images using synthetic point spread functions generated for each filter using the TinyTim program (Krist & Hook 1999), as there are no bright point sources within the PC2 chip, but this did not noticeably improve the image resolution, and in no case (or particular iteration of the deconvolution process) did the relative brightnesses of the different peaks change. This could arise from a mismatch of the synthetic and actual point spread functions, but we instead suspect errors in the deconvolution process used by D2000. We thus restrict our analysis to the properties of the central brightness peak, given its dominance of the galaxy’s emission.

Our images, notably the deep F814W image, confirm POX 186 to be extremely small. Its maximum angular size in the F814W image is approximately 3″, corresponding to only 300 pc at our adopted distance. The central brightness peak appears unresolved in the F336W image, and marginally resolved in the F555W and F814W images, with a small elongation in the east-west direction. This indicates its size to be ~10 - 15 pc. We will refer to this region as the central cluster, although there are indications of subcomponents within it.

A striking feature of the contour plot of the F336W image is the detection of what appear to be individual or small groups of stars arrayed in an arc around the center of the galaxy. The F555W and F814W plots show this arc to partially overlap with the tail of material on the object's western side, suggesting a longer loop or stream too faint to be detected in the F555W and F814W images. Another feature evident in the contour plots of the F555W and F814W images is the compression of the isophotes immediately to the west of the central cluster, indicating a strong stellar density gradient in that direction.

We obtained aperture photometry of the three separate images using the IRAF "polyphot" task, which allows polygonal apertures to be custom fit around objects. This is particularly appropriate to POX 186 given its irregularity. The resulting apertures have a maximum size slightly exceeding the galaxy's size of 3″. The resulting magnitudes were placed on a Vega scale using the zeropoints given by Whitmore (1995). These magnitudes are given in Table 2, along with the corresponding fluxes. At the adopted distance to the galaxy the F555W flux corresponds to a luminosity ~$10^{-4}$ $L$*, modulo contamination from the nebular lines included in this filter (principally [O III] λλ4959, 5007; the correction to the flux value given in Table 2 should be approximately 0.3 dex).

The red color of the regions beyond the central and secondary brightness peaks (Figure 1) is consistent with the findings of D2000. To examine this quantitatively, we subtracted the F814W image from the F555W image and measured the radial color profile, averaged across the entire galaxy (the two-dimensional difference image has a high level of pixel-to-pixel variance, necessitating an average). The result is shown in Figure 3. The colors of the regions approximately 1.2″ from the central star cluster fall in the range of late K / early M stars (both main sequence and giant). Contamination of the stellar continuum by nebular lines, particularly [O III] λλ4959, 5007 should not be very strong that far away from the central cluster, given its compactness (Figure 2). The same plot for the F334W – F555W difference image is noise-dominated at similar distances, so this color cannot be used as a further diagnostic of the stellar population.

## 3.2 Spectra

### 3.2.1 Emission-Line Analysis

The combined ultraviolet and optical spectrum of POX 186 is shown in Figure 4. The gap between approximately 1670 Å and 1770 Å is due to trimming of the G230L spectrum at wavelengths where it became noise-dominated. The flux levels of both STIS spectra were lower than expected on the basis of scaling the optical fluxes and the assumption made about the form of the UV continuum, which resulted in lower signal-to-noise ratios than expected. This may be due in part to dust extinction in the galaxy, for which we find evidence on the basis of the Balmer line ratios, discussed further below. We detect Lyα in emission and absorption, but do not attempt to measure the strength of these features due to the large uncertainties. Emission in C III] λ1909 is also detected, but is similarly noisy. Blended lines in the optical spectrum were measured after Gaussian fitting. Certain values, in particular, those for the [S II] λλ6714, 6731 lines, are uncertain due to blending and weakness. Kunth & Joubert (1985) find "suggestive" evidence of a broad He II λ4686 feature indicative of Wolf-Rayet stars, which we do not confirm in our spectrum, although there is a suggestion of weak broad He II λ1640 emission in the STIS spectrum.

In order to estimate the electron density $n_e$, electron temperature $T_e$, and reddening E($B$ – $V$) of the ionized gas, a set of five-level model atoms were constructed from which optical line ratios could be calculated as a function of density and temperature. We then solved simultaneously for the $n_e$, $T_e$ and E($B$ – $V$) values that reproduce three observed line ratios: Hα/Hβ or Hγ/Hβ, [O III] λ4363 / [O III] λ4959, and [A IV] λ4711 / [A IV] λ4740 or [S II] λ6716 / [S II] λ6731. Uncertainties on the parameter values were determined using a Monte Carlo procedure that uses errors on the observed line fluxes to generate large sets of "synthetic" flux ratios, and then solves for the parameters of each set. From the [A IV] λ4711 / [A IV] λ4740 and [S II] λ6716 / [S II] λ6731 ratios we find the density to be near the low-density limit, $n_e$ ~100 $cm^{-3}$, to within the uncertainties in the line fluxes. Assuming such a density, the Balmer line ratios indicate $T_e \cong 15{,}900 \pm 700$ K and E($B$ – $V$) $\cong 0.28 \pm 0.01$. This reddening value is in good agreement with that of Kunth & Sargent (1983).

The ratios of the optical line fluxes to that of Hβ are given in Table 3. These have been measured from both the observed frame spectrum of POX 186 after correction for Galactic extinction, and from the rest-frame spectrum after applying the above internal reddening correction using the extinction law of Calzetti, Kinney & Storchi-Bergmann (1994). These measurements yield a limit on the ratio of oxygen to hydrogen of 12 + log(O/H) > 7.73. A stronger constraint is not possible due to the lack of a clear detection of any O II lines such as [O II] λ7325 ([O II] λ3727 was not covered in our observed wavelength range). Nevertheless, the limit we derive for the O/H abundance ratio agrees to within 10% with the value reported by Pagel et al (1992), as derived from the data of Kunth & Sargent (1983)[2]. If we assume that the metallicity scales linearly with the oxygen abundance, we find a metallicity for POX 186 of $Z \cong 0.06\ Z_{\odot}$, assuming log $(O/H)_{\odot}$ = -3.08 (Anders & Grevasse 1989).

### 3.2.2 Properties of the Central Cluster

Although our optical spectrum includes emission from all of POX 186, it is dominated by the central cluster, particularly its OB stars. This can be seen in Figure 4 by the continuity between the UV and optical continua. The optical spectrum also shows no clear absorption features, and an estimate of the number of O stars in the central cluster (given below) yields an optical flux level very similar to what we measure. We are still able to estimate the age of this cluster, if not the underlying late-type stars in the galaxy, in two ways. First, comparison of the observed Hβ equivalent width (Table 3) with the predictions of the Starburst99 models (Leitherer et al. 1999) for the case of a low-metallicity instantaneous starbursts with a Salpeter initial mass function and an upper mass limit of 100 $M_{\odot}$ yields an age of approximately 3 Myr (an age of approximately 6 Myr is inferred for a continuous burst). Second, we fit the optical continuum with population synthesis models from the library of Bruzual & Charlot (2002). We constructed continuum models of single stellar populations (instantaneous bursts) of various ages, assuming a Salpeter initial mass function spanning the range 0.1 $M_\odot$ to 100 $M_\odot$ and a metallicity of 0.2 $Z_\odot$ (the value in the Bruzual & Charlot 2002 library closest to our estimate for the metallicity of POX 186). The model continua were fit to the observed spectrum over the range 4200 Å - 7300 Å, excluding emission lines and using a least-squares routine in which the reddening was allowed to vary. The best fitting model was found to have an age of 4 Myr (+ 1, -1.6 Myr), and a reddening of E($B - V$) ≅ 0.3 (± 0.15), using the Calzetti et al. (1994) extinction law. This reddening value is consistent with that obtained from our emission-line analysis (§ 3.2.1). The resulting fit is shown in Figure 5. From this best-fitting model we also infer a mass to the cluster of approximately $10^5 M_\odot$, qualifying it as a "super" star cluster of the type seen in other BCDGs, including Henize 2-10 (see Johnson et al. 2000 and references therein). Comparable results are obtained if the ultraviolet spectra are included in this fit, but more uncertainty is introduced by their higher noise level and discontinuity.

[2] Kunth & Sargent (1983) note contamination by second-order light in their spectra at wavelengths longer than 6300 Å. They thus simply assumed an Hα/Hβ ratio of 2.8, and scaled the line fluxes in the red part of their spectra accordingly. Since they performed this scaling before applying internal reddening corrections, abundances derived from their spectra redward of 6300 Å may be incorrect.

Several other important properties of the central cluster can be measured from the dereddened spectrum. First, the total number of photons in the Lyman continuum can be estimated from the Hβ luminosity (Osterbrock 1989). From the temperature estimated above and the dereddened Hβ luminosity we find $N$_LyC = $6.59 \times 10^{51}$ s$^{-1}$. Assuming log($N$_LyC) = 49.12 s$^{-1}$ for a single O7V star (Vacca, Garmany & Shull 1996), the tables of Vacca (1994) with corrections for age provided by Schaerer (1996) indicate ~1000 – 2000 equivalent O V stars in the central cluster for the object's measured metallicity for a Salpeter initial mass function and an age of 3-4 Myr. This range of values for the number of stars may however only be a lower limit if the ionized gas is density bounded (leading to an escape of more Lyman continuum photons), as has been found in the Wolf-Rayet galaxy NGC 4214 (Leitherer et al. 1996) after spatially resolving its star-forming regions. The star formation rate of the cluster can be estimated from the strength of the UV continuum using the scaling relation of Madau, Pozzetti & Dickinson (1998): SFR ($M_\odot$ yr$^{-1}$) = $1.25 \times 10^{-28} L_{1500}$, where $L_{1500}$ is the rest-frame continuum luminosity at 1500 Å in ergs s$^{-1}$ Hz$^{-1}$. Measuring $L_{1500}$ from the STIS spectrum (corrected for internal extinction using our measured E($B - V$) value) and our adopted distance yields SFR = 0.045(±0.003) $M_\odot$ yr$^{-1}$. Measurement of the SFR using the Kennicutt (1983) relation for Hα yields a higher rate of 0.08(±0.002) $M_\odot$ yr$^{-1}$. However, Kennicutt (1992) notes that the SFRs of BCDGs may be up to 3 times lower than those yielded by the Kennicutt (1983) calibration, which was derived from spiral galaxies, so a correction by this factor would bring these estimates into better agreement. These star formation rate values are also only upper limits to the true star formation rate of the cluster, given that both the Madau et al. (1998) and Kennicutt (1983) relations strictly apply to systems forming stars continuously over timescales greater than ~$10^8$ years, whereas our results indicate that the stars in the central cluster of POX 186 have been formed in a very recent (~$10^{6-7}$ yr) burst.

### 3.3 Environment

The environment of POX 186 is relevant to the question of its origin. A search of its surroundings from both the Digitized Sky Survey plates and using the NASA Extragalactic Database reveals no galaxies to within roughly 5 Mpc of it. Thus while there is good evidence that some dwarf galaxies form out of the debris produced by the collision of spiral galaxies (e.g. Braine et al. 2000; Weilbacher et al. 2000; Hunter, Hunsberger & Roye 2000), such a model cannot explain POX 186. This also argues against a model in which its asymmetric morphology was produced by an encounter with a more massive galaxy. Under the assumption that, as argued in § 3.1, the objects detected in the F814W image are background sources, this indicates that POX 186 resides in a void. Indeed, the galaxy's position ($\alpha$ = 13h 25m, $\delta$ = -11° 36′) suggests that it lies on the periphery of the Boötes void (e.g. Kirshner et al. 1987).

## 4. DISCUSSION

POX 186 is clearly a very small, dynamically disturbed object. The tail of stars seen in the F555W and F814W images, and which the F336W image suggests is a part of a longer stream of material, is qualitatively similar to features seen in interacting disk galaxies (e.g. Hibbard & van Gorkom 1996) which are likely due to tidal effects (e.g., Toomre & Toomre 1972). In combination with its asymmetry and tail, the compression of the isophotes to the west of the central star cluster evident in the F555W and F814W contour plots suggests that POX 186 represents a recent collision between two small clumps of stars of comparable mass and size (~100 pc). Given that there are no nearby massive galaxies with which it might have interacted (§ 3.3), this interpretation seems preferable to one in which POX 186 was originally a single, dynamically relaxed object. More specifically, the history suggested by the contour plots is that a less massive clump spiraled into a larger clump from a direction due east of the object's center, leaving a tidal stream of material behind in which star formation was triggered. The less massive clump then impacted the larger clump on its northwestern side, producing a steep stellar density gradient in that direction as well as the central starburst. A timescale for this encounter of less than ~$10^8$ yr ago is suggested by the presence of early-type stars in the putative tidal tail and arc (Figure 2) and by the young age of the central star cluster.

This "colliding clumps" hypothesis can be checked by a simple estimate of the dynamical timescale of such a system. Specifically, we can test whether a collision could have occurred within the timescale of $10^8$ yr suggested above. Our estimate of $L \sim 10^{-4} L^*$ for POX 186 corresponds to a $B$-band luminosity of approximately $2.7 \times 10^6 L_{\odot}(B)$. Thuan (1987) argues that the ratios of total masses to $B$-band luminosities of BCDGs are in the range ~ 2 – 4, which yields masses of approximately $5 \times 10^6 M_{\odot}$ to $10^7$ $M_{\odot}$ for POX 186 for this luminosity. Kunth et al. (1988) also estimate a mass ~$10^7 M_{\odot}$ for POX 186 from their optical photometry. Considering only their gravitational attraction, clumps of mass ~$5 \times 10^6 M_{\odot}$ and size ~100 pc that are just now in contact would have been separated by approximately 400 pc at a time $10^8$ years ago, assuming radial motion. It is therefore possible that POX 186's morphology and current star formation are the results of such a collision. A better check of a collision model must await information on the galaxy's kinematics. Deep, high-resolution 21 cm line observations are particularly important. Interestingly, H I imaging of the environments of nearby irregular galaxies by Wilcots, Lehman & Bryan (1996) has revealed them to have surrounding H I clouds with masses that are also ~$10^7 M_{\odot}$. These authors argue that such clouds are in the process of being accreted onto the primary galaxy as part of an ongoing formation process, similar to what we have proposed for POX 186.

If the above interpretation of POX 186's morphology is correct, it suggests that the object represents a very small dwarf galaxy that is still in the process of formation. Clumps of stars ~100 pc in size and ~$10^{6\text{-}7}$ $M_{\odot}$ that are not obviously globular clusters could be the long sought-after galaxy building blocks required by hierarchical models of galaxy formation (e.g. Cole et al. 2000 and references therein). The oldest stars in POX 186 may also be just a few billion years old, based on our results and on age estimates for BCDGs with morphologies similar to POX 186, including Mrk 59 and Mrk 71 (Noeske et al. 2000) and Tol 1214-277 (Fricke et al. 2001). Raimann et al. (2000) obtain similar results for a larger sample of BCDGs. It is unfortunately not possible to make an age estimate for POX 186's oldest stars with the available data, as even our PC2 images do not adequately resolve individual stars in the galaxy, and the OB stars of the central star cluster dominate our spectra. A spectrum away from the central star cluster is needed for this, which because of the galaxy's small angular size and faintness would best be obtained with an 8m-class telescope under excellent seeing. The evidence for internal dust and the galaxy's non-zero metallicity establishes that there have been previous episode(s) of star formation, however, which would appear to have occurred within the separate clumps of stars before their collision. An interesting question to consider is the effect that the eventual supernovae in the central star cluster will have on the subsequent evolution of the galaxy, both in terms of stellar dynamics and star formation rate. A significant impact on both parameters seems likely, given the relatively large fraction of the galaxy's mass contained in the cluster (~0.01).

A small number of apparently ultracompact (~100 pc) but quiescent objects have recently been identified in the core of the Fornax cluster (Drinkwater et al. 2000; Phillipps et al. 2001). However, the nature of these objects, in particular whether they represent only the cores of very low surface brightness galaxies, remains highly uncertain. Comparisons between these objects and BCDGs would thus be premature, especially in the case of POX 186, which resides in a void rather than a cluster (§ 3.3). The isolation of POX 186 is in fact typical of a large fraction of the BCDG population (Campos-Aguilar, Moles & Masegosa 1993; Popescu, Hopp & Rosa 1999). Interestingly, a number of larger galaxies in the Boötes void also show evidence for dynamical disturbance and recent interaction (Cruzen, Weistrop & Hoopes 1997; Cruzen et al. 2002) similar to POX 186. These results suggest that such voids are the last places that galaxies form. The formation of dwarf galaxies in voids is also consistent with the most basic Cold Dark Matter cosmologies (Dekel & Silk 1986).

Finally, if POX 186 is indeed still forming, it would support the "downsizing" picture of galaxy formation first proposed by Cowie et al. (1996), in which the least massive galaxies are the last to form.

The results of Noeske et al. (2000) and Raimann et al. (2000) that BCDGs in the local universe may only be a few billion years old support such a model. The inverse correlation between galaxy mass and mass-normalized or "specific" star formation rate found by Guzmán et al. (1997) and Brinchmann & Ellis (2000) has also been taken as evidence for the downsizing picture, and Kruger, Fritze-von Alvensleben & Loose (1995) note that BCDGs show a similar trend of stronger starbursts for lower masses. Our estimate of POX 186's star-formation rate along with its estimated mass of $10^7\,M_\odot$ yields a specific star formation rate of log (SFR/ $M_\odot$ ) = -8.3. This value is roughly consistent with a visual extrapolation of the correlations found by Guzmán et al. (1997) and Brinchmann & Ellis (2000) to a galaxy mass of $10^7\,M_\odot$ (noting that we have necessarily used different estimators for mass and star formation rate than those studies). In addition to POX 186's morphology, this supports the idea that these correlations are the result of a downsizing process. Our results also suggest that the star formation of BCDGs is being driven by the merging of sub-galactic clumps, with POX 186 being a relatively obvious case. Statistical evidence for merger-driven star formation among BCDGs has recently been presented by Noeske et al. (2001). High-resolution imaging and kinematic studies of more "ultracompact" blue dwarf galaxies are required confirm this.

We thank Jarle Brinchmann, Nelson Caldwell, Stephane Charlot, Don Garnett, John Hibbard, Ed Olszewski and Glenn Schneider for helpful discussions and assistance with these data, and an anonymous referee for helpful comments. This work was supported by Computer Sciences Corporation, and through NASA by grant number GO-08333.01-97A to the Space Telescope Science Institute. The Space Telescope Science Institute is operated by the Association of Universities for Research in Astronomy, Inc., under NASA contract NAS5-26555. This work has also made use of the NASA Extragalactic Database, which is operated by the Jet Propulsion Laboratory under contract with NASA.

TABLE 1

HUBBLE SPACE TELESCOPE OBSERVATIONS OF POX 186

| Instrument | File Name | Filter | UT Date | Integration Time, s |
|---|---|---|---|---|
| WFPC2 | U5GB0101R | F336W | 3/15/2000 | 700 |
| WFPC2 | U5GB0102R | F336W | 3/15/2000 | 700 |
| WFPC2 | U5GB0103R | F555W | 3/15/2000 | 600 |
| WFPC2 | U5GB0104R | F814W | 3/15/2000 | 2700 |
| WFPC2 | U5GB0105R | F814W | 3/15/2000 | 2700 |
| STIS | O5GB02010 | G140L | 6/16/2000 | 2100 |
| STIS | O5GB02020 | G230L | 6/16/2000 | 2082 |

TABLE 2

POX 186 APERTURE PHOTOMETRY

| Filter | Magnitude (Vega scale) | Log Flux, ergs $s^{-1}$ $cm^{-2}$ $Å^{-1}$ |
|---|---|---|
| F336W | 17.4 (± 0.1) | -15.44 (± 0.02) |
| F555W | 17.6 (± 0.1) | -15.48 (± 0.01) |
| F814W | 18.4 (± 0.1) | -16.28 (± 0.01) |

TABLE 3

POX 186 EMISSION LINE RATIOS[1]

| Line | $f_\lambda$(line)/ $f_\lambda$(Hβ) (1) | $f_\lambda$(line)/ $f_\lambda$(Hβ) (2) |
|---|---|---|
| Hγ | 0.43 (± 0.01) | 0.49(± 0.01) |
| [O III] λ4363 | 0.11(± 0.01) | 0.12(± 0.01) |
| He I λ4471 | 0.035(± 0.006) | 0.038(± 0.006) |
| [Ar IV] λ4711 | 0.025(± 0.004) | 0.025(± 0.004) |
| [Ar IV] λ4740 | 0.019(± 0.006) | 0.020(± 0.006) |
| [O III] λ4959 | 1.985(± 0.004) | 1.943(± 0.004) |
| [O III] λ5007 | 5.908(± 0.002) | 5.723(± 0.002) |
| He I λ5876 | 0.095(± 0.003) | 0.078(± 0.003) |
| [O I] λ6300 | 0.021(± 0.004) | 0.016(± 0.004) |
| [S III] λ6312 | 0.016(± 0.001) | 0.012(± 0.001) |
| Hα + [N II] λλ 6548,6583 | 3.776(± 0.008) | 2.780(± 0.008) |
| [S II] λ6716 | 0.060(± 0.001) | 0.043(± 0.001) |
| [S II] λ6731 | 0.041(± 0.002) | 0.029(± 0.002) |
| He I λ7065 | 0.046(± 0.002) | 0.035(± 0.002) |
| [A III] λ7136 | 0.083(± 0.002) | 0.057(± 0.002) |
| [A III] λ7751 | 0.019(± 0.003) | 0.012(± 0.003) |
| $f_\lambda$(Hβ) (ergs $s^{-1}$ $cm^{-2}$ ) | 2.31(± 0.02) $\times 10^{-14}$ | 5.88(± 0.04) $\times 10^{-14}$ |
| EW(Hβ)(Å) | 262 (± 21) | |

TABLE 3 (CONTINUED)

---

[1] Column (1) values are for spectrum in observed frame, after correction for Galactic extinction. Column (2) values are in the object rest frame, after correction for an internal reddening of E(*B*-*V*) = 0.282.

# FIGURE CAPTIONS

FIG. 1. -- Color composite of the F336W, F555W and F814W Planetary Camera 2 images of POX 186, where the images have been placed in the blue, green and red channels, respectively.

FIG. 2. -- Contour plots of the Planetary Camera 2 images of POX 186. The image orientation is the same as in Figure 1. The images are flux calibrated, with the following minimum and maximum contours in log (ergs $s^{-1}$ $cm^{-2}$ $Å^{-1}$ ): -16.34 to -13.94 (F336W), -16.60 to -14.90 (F555W) and -16.25 to -14.55 (F814W). The contour intervals are 0.2 dex for the F336W image, and 0.1 dex for the F555W and F814W images. At the adopted distance to the galaxy of 21 Mpc, the angular scale is approximately 100 pc $arcsec^{-1}$ .

FIG. 3. -- Profile of the F555W – F814W difference image, averaged over the entire galaxy. The vertical axis is in units of Vega magnitude, and the horizontal axis has the central brightness peak as its origin.

FIG. 4. -- Combined STIS ultraviolet and ground-based optical spectra of POX 186, in the observed frame of the galaxy, after correction for Galactic extinction. The gap in the STIS spectra is due to trimming of the G140L and G230L grating spectra in regions of very low signal-to-noise levels.

FIG. 5 -- Optical spectrum of POX 186 in the object rest frame, with the best-fitting Bruzual & Charlot (2002) population synthesis model shown (bold line; see text for details).